\documentclass{iopjournal}
\usepackage{float}
\usepackage{graphicx}   
\usepackage{subcaption} 

\begin{document}

\articletype{Paper} 

\title{Analysis and Prediction of Dark Current Mechanisms in Si:P Blocked Impurity Band (BIB) Infrared Detectors}

\author{Mengyang Cui$^{1,2,3}$\orcid{0009-0000-2396-3932}, Hongxing Qi$^{1,2,3}$\orcid{0009-0005-1762-4607}, Chengduo Hu$^{1,2,3}\orcid{0009-0005-8681-1924}$ and Qing Li$^{1,2,3*}$\orcid{0000-0000-0000-0000}}

\affil{$^1$School of Physics and Optoelectronic Engineering, Hangzhou Institute for Advanced Study, Hangzhou, China}

\affil{$^2$Shanghai Tnstitute of Technical Physics, Chinese Academy of Science, Shanghai, China}

\affil{$^3$University of Chinese Academy of Sciences, Beijing, China}

\affil{$^*$Author to whom any correspondence should be addressed.}

\email{liqing@ucas.ac.cn}


\begin{abstract}
We investigated the nonlinear phenomena observed in the dark current of BIB (blocked-impurity-band) 
infrared detectors, including negative differential resistance (NDR) and current oscillations. Our analysis systematically elucidated the intrinsic transport mechanisms in optimized devices, revealing that these anomalies arise from current path clustering induced by structural disorder and impurity band conduction dynamics. Notably, the simulated current-voltage (I-V) characteristics demonstrated strong agreement with experimental measurements across a wide bias range, confirming the validity of our proposed physical model.Furthermore, we developed a transformer-based predictive model using experimental dark current datasets. The model achieved robust performance metrics and this framework enables rapid prediction of dark current trends under varying operational conditions, providing actionable insights for detector optimization. 
\end{abstract}

\section{Introduction}
Silicon-based blocked-impurity-band (BIB) infrared detectors, particularly Si:P BIB detectors, have emerged as 
critical components in mid- to far-infrared (MWIR/FWIR) sensing applications. These devices exploit the unique electronic properties of heavily doped silicon layers to detect radiation in the 35-75 $\mu m$ spectral range, a region of interest for astrophysical observations, planetary science, and environmental monitoring. Unlike conventional extrinsic photodetectors (e.g., HgCdTe), Si:P BIB detectors utilize impurity band conduction mechanisms, where charge carriers are collected from both the conduction band and impurity band states. This dual-carrier collection mechanism reduces recombination noise, enabling high quantum efficiency ($>50\%$) and low dark current ($<1 nA/cm²$) under cryogenic conditions. The extended cutoff wavelength of Si:P BIB detectors (up to 75 $\mu m$) positions them as indispensable tools for studying cosmic phenomena, such as interstellar dust emission and thermal radiation from cool celestial objects\cite{1}. For instance, NASA's James Webb Space Telescope (JWST) employs Si:P BIB detectors in its Near-Infrared Spectrograph (NIRSpec) to characterize exoplanetary atmospheres in the 2-5 $\mu m$ range. Additionally, their compatibility with CMOS readout circuits and radiation hardness make them suitable for space missions requiring long-term operational reliability. 
BIB infrared detectors hold significant application value in advanced sensing technologies, however, their 
performance remains severely constrained by excessive dark current\cite{2,3}. Current optimization strategies predominantly rely on iterative adjustments of device parameters followed by dark current measurements, a labor-intensive process hindered by the absence of robust physical models\cite{4}. Conventional simulation tools such as Sentaurus TCAD and VASP struggle to achieve consistency with experimental results due to incomplete descriptions of carrier transport mechanisms in these devices. To address this challenge, there is an urgent need for updated physical models that can elucidate the underlying charge transport phenomena, particularly in heavily doped semiconductor layers and impurity band conduction regimes. Integrating existing experimental datasets, such as temperature-dependent dark current measurements and bias-voltage response profiles, into model calibration could accelerate the development of predictive frameworks. Furthermore, systematic analysis of experimental data may reveal critical insights into material defects, impurity band dynamics, and interfacial charge transfer processes, thereby guiding targeted device architectures to mitigate dark current contributions while preserving quantum efficiency. 


\begin{figure}
 \centering
        \includegraphics[width=0.5\textwidth]{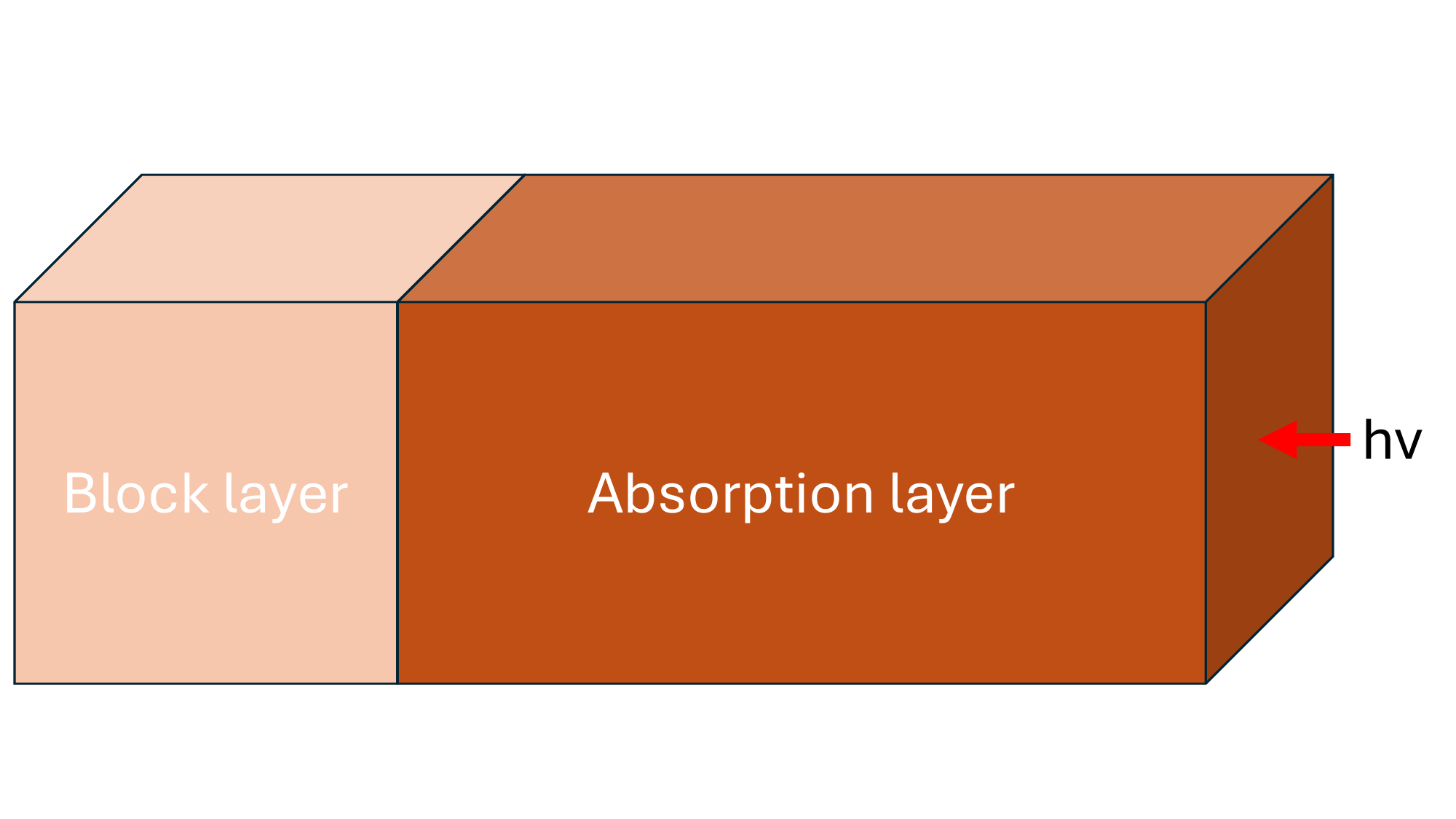}
 \caption{A simple model of Si:P BIB which is a cuboid silicon doped with phosphorus. The dopping design is at the concentrations of lower than $1\times 10^{15}cm^{-3}$ at block layer and $1 \times 10^{18}cm^{-3}$at absorption layer. 
 The length of the blocking layer is typically around 5 $\mu m$, the absorbing layer is typically around 20 $\mu m$, and the width of the cross-section can be adjusted in the range from several micrometers to 200 $\mu m$.}
\label{fig1}
\end{figure}

\section{Model and Experiments}
\subsection{Dark current mechanism}
Our research strategy prioritizes the investigation of anomalous experimental data to elucidate underlying physical 
mechanisms, followed by the analysis of typical data patterns. This approach is grounded in the hypothesis that extreme values observed in special datasets inherently reflect the intrinsic physical mechanisms operative within 
standard device architectures. The BIB device exhibits remarkably pronounced nonlinear 
current phenomena, which demonstrates the localization phenomenon induced by doping\cite{5}. 


\begin{figure}[H]  
 \centering
        \includegraphics[width=0.9\textwidth]{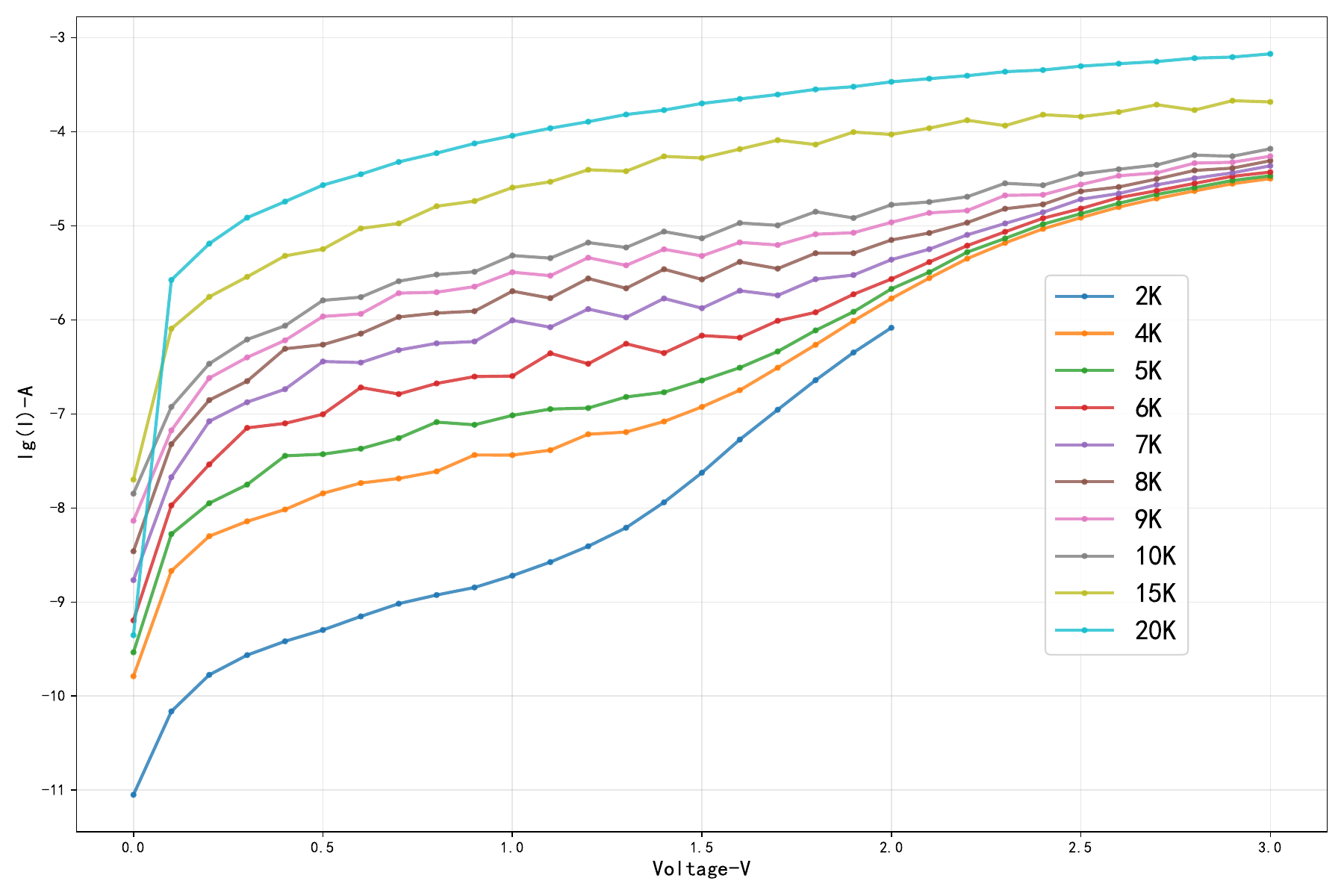}
 \caption{A small number of devices exhibit dark current oscillation phenomena, which may indicate particularly 
inhomogeneous doping of the silicon in this batch. Such distinctive dark currents, manifested under particularly prominent conditions, suggest that they also exist in ordinary situations but are just less pronounced. The unit of the ordinate is the logarithm of the dark current I to the base 10 (i.e., lg(I)), with the unit of ampere (A). The abscissa represents the applied voltage of the device, with the unit of volt (V).}
\label{fig2}
\end{figure}

The electrical transport properties of BIB devices are particularly influenced by doping \cite{6,7}. For this 
reason, we need to analytically construct physical models to analyze their nonlinear phenomena\cite{8,9,10}. According to Aladashvili's 
theoretical model where the device can be abstracted as a series-connected structure of distinct regions\cite{11,12}. The observed oscillatory behavior is attributed to the extremely non-uniform distribution of phosphorus atoms within the lightly doped regions of the blocking layer. According to the model, the primary charge carriers (electrons injected from electrodes) exhibit complex transport dynamics through these lightly doped regions. Specifically, their movement trajectories form an interconnected network dominated by phosphorus ion pathways, characterized by multiple primary channels and secondary branches that create preferential conduction paths. This structural anisotropy in the dopant distribution leads to spatially modulated \cite{13,14}, which manifests as measurable oscillations in device characteristics. 

\begin{figure}[H]  
 \centering
        \includegraphics[width=0.9\textwidth]{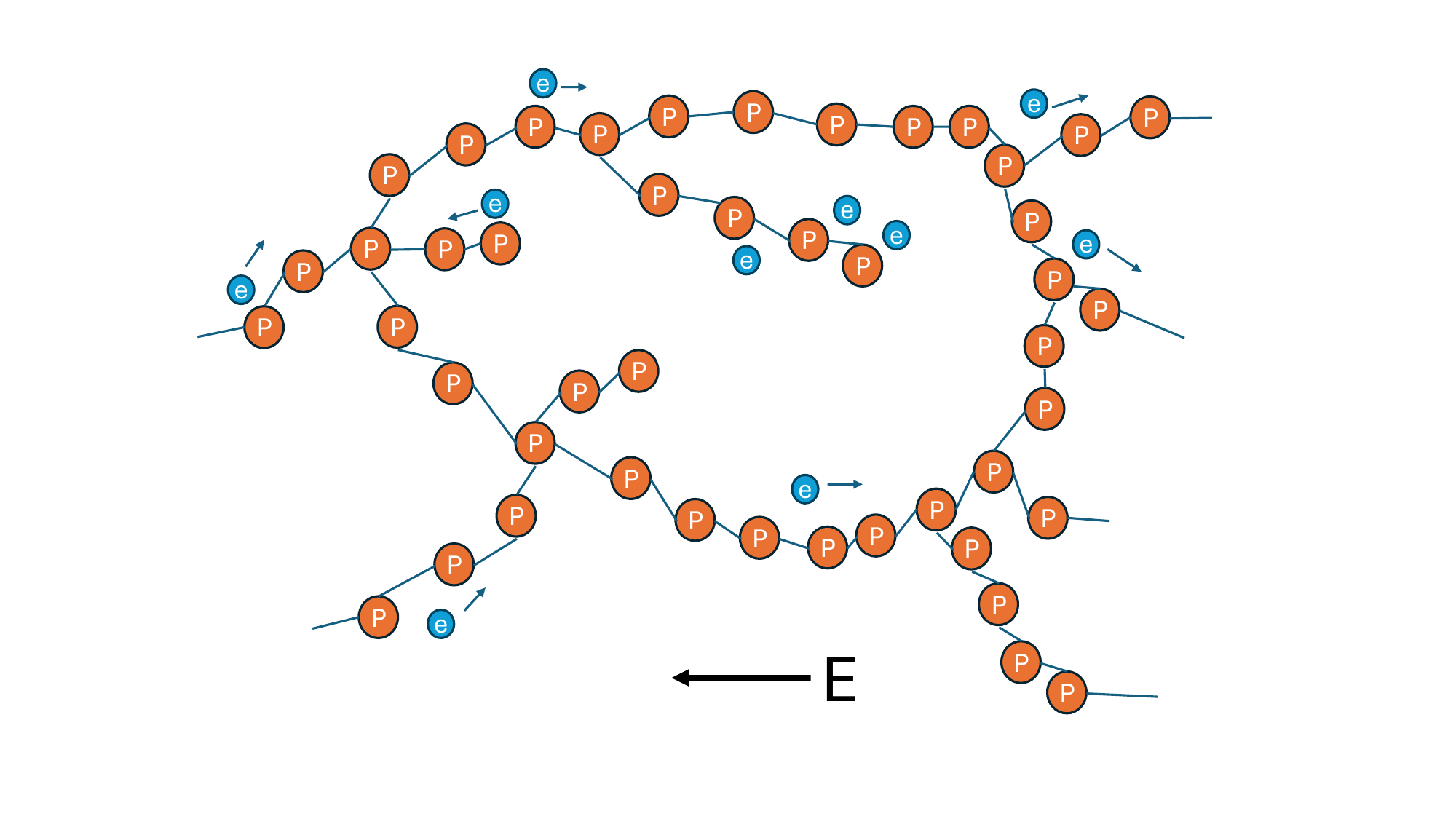}
 \caption{Microscale current network. The figure omits the background silicon atoms and only highlights the local current paths and partial current blocking phenomena.
 The orange spheres in the figure represent phosphorus atoms. The black lines denote the paths, which may also pass through silicon during the hopping process of phosphorus atoms, however, this physical process is omitted in the figure.}
\label{fig3}
\end{figure}

We define the characteristic length of such clusters as $ L_0$. When $e E L_0\ll \mathcal{K}T$  , electrons can freely hop between different pathways. However, when $e E L_0\gg  \mathcal{K}T$  , some electrons in the branch cannot return to the previous node against the direction of the electric field, nor do they have sufficient energy to hop to the next lattice site. Consequently, they get trapped in the branch, leading to an increase in resistance. However, as local electrons accumulate, some electrons are enabled to perform long-distance hops to the next lattice site, which results in current oscillations. This persists until the field strength increases further, at which point all electrons are driven by the electric field to drift rather than hop between lattice sites. Macroscopically, this manifests as oscillations starting at a certain voltage and then gradually diminishing.
A rough description of the resistivity variation induced by this mechanism is given by the formula $$\sigma(E)\approx \sigma(0) \exp (-\frac{e E L_0 }{2 k T})$$

Given that the theoretical framework is in good agreement with the experimental phenomena, and that charge conduction is assumed to be hopping conduction, this current oscillation phenomenon can thus be utilized to determine the components of the current conduction present and the doping uniformity.
It can be observed from Fig. 4 that even under low-temperature conditions, current oscillations still occur in the device at an applied bias voltage of approximately 5 V. Furthermore, an increase in temperature leads to a decrease in the onset voltage of these oscillations. This is likely because this type of device is highly temperature-sensitive, and a portion of thermally excited carriers also participate in conduction via hopping current. For devices with highly non-uniform doping, when current oscillations initiate, the oscillations may even intensify as the voltage increases. This phenomenon reflects the characteristics of a typical nonlinear system, indicating both the continuous variation of current flow paths and the severe non-uniformity of doping.
Thus far, conclusions have been drawn from a limited set of atypical data that the source of dark current in BIB devices is highly correlated with hopping current, and that doping and temperature exert a significant influence on the characteristics of the current\cite{15}.
\begin{figure}[H]  
 \centering
        \includegraphics[width=0.9\textwidth]{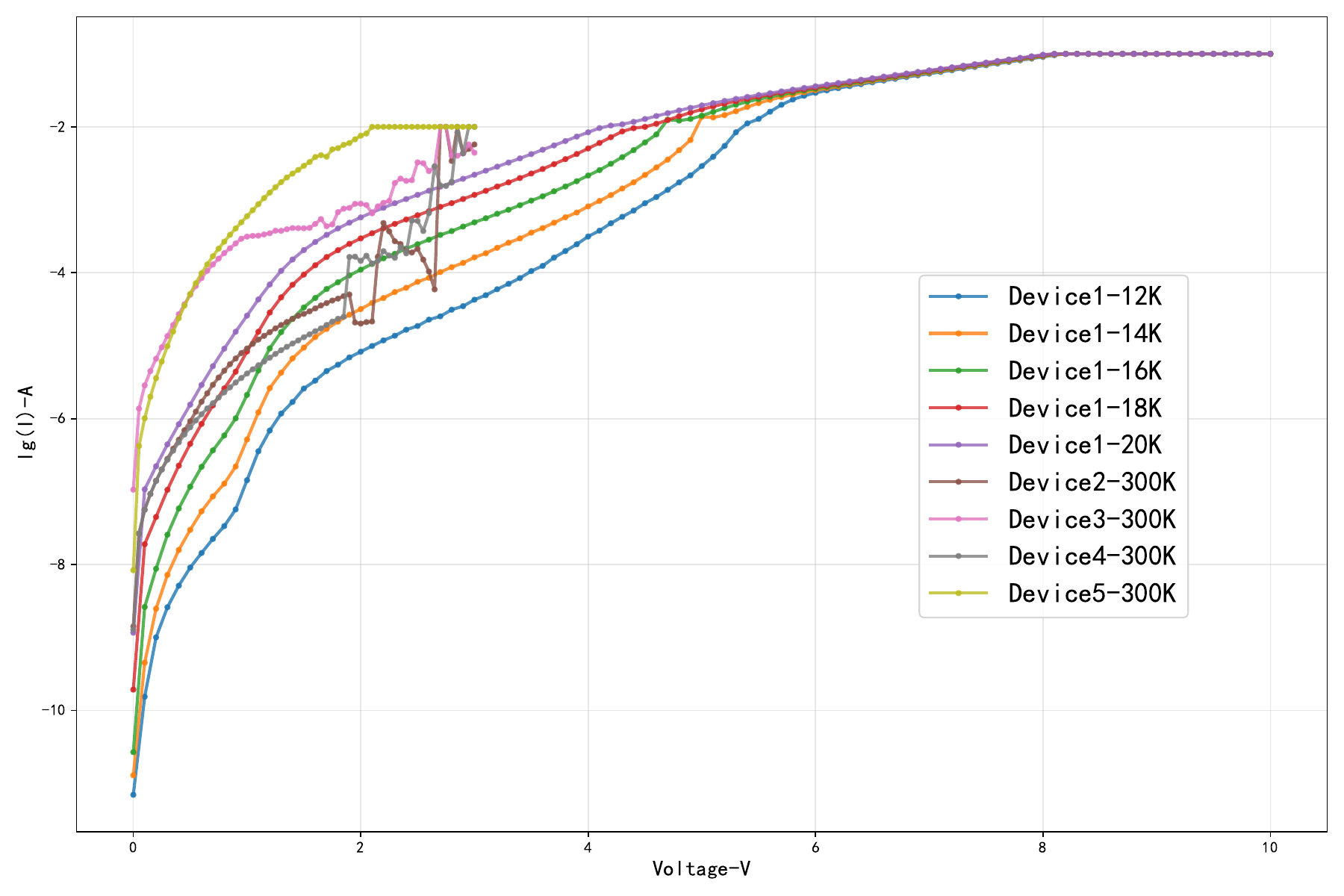}
 \caption{Dataset exhibiting negative differential resistance phenomena in dark current measurements.}
\label{fig4}
\end{figure}
Next, we consider the more general case where current oscillations are not observed. We assume that the doping is relatively uniform under such circumstances, and there are no significant lattice defects at the interface. Even if minor current oscillations exist in this scenario, they are filtered out by dark current measurement instruments.
We employ the space charge limited currents (SCLC) theory to elucidate the intrinsic transport characteristics of doped silicon under such conditions\cite{16,17,18}. 
The SCLC theory posits that the electric field-dependent current characteristics in materials are predominantly governed by the interplay between dopant-induced disorder and Frenkel localization effects. These mechanisms synergistically induce power-law scaling in current-voltage (I-V) curves. In devices with significant trap state densities (far exceeding dopant atom concentrations), charge carriers exhibit a hierarchical occupation process: low-energy traps are preferentially filled before deeper energy levels become accessible. This sequential charge injection dynamics manifests as distinct slope variations in the I-V characteristics, reflecting the energy-dependent trapping and detrapping kinetics\cite{19}. 

\subsection{Experiments}

We have accumulated measured data of dark currents for devices with various designs. Measurements on different 
devices are concentrated in the voltage range of -10V to 10V, generating tens of thousands of data points, which have been constructed into the dataset required for training. 
Additionally, we developed a program in Python to simulate the electronic transport properties of silicon with relatively uniform phosphorus doping and calculate the corresponding current surface density.
\begin{figure}[H]  
 \centering
        \includegraphics[width=1\textwidth]{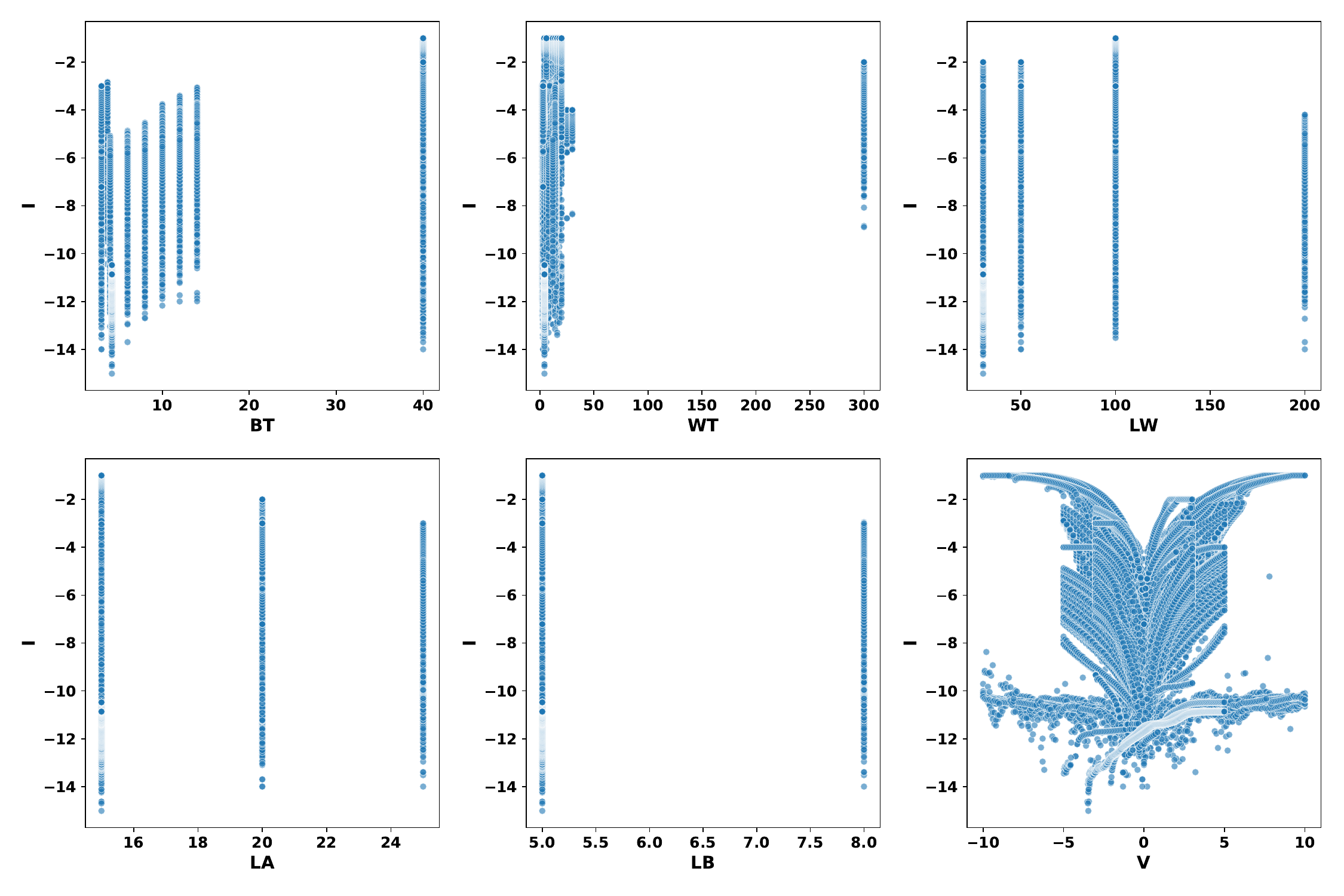}
 \caption{Overview of Data Distribution Characteristics.The current I here refers to lg(I), with the unit of ampere (A). Both WT and BT have the unit of absolute temperature, Kelvin (K). LB, LW, and LA have the unit of micrometer ($\mu m$). The voltage V has the unit of volt (V).}
\label{fig5}
\end{figure}
We build a lightweight Transformer-based regression framework for current prediction consists of several core components working in sequence\cite{20}. 
The framework using six input 
features (BT, WT, LW, LA, LB, V) with the target being the current (I), where WT means the working temperature, LA means the length of absorption layer, the V means the applied voltage, the BL means the length of block layer, and the LW is the length of the side length of the square photosensitive window and BT is the background temperature. The data processing pipeline begins with separating features 
and the target variable from the dataset, followed by scaling using RobustScaler to handle 
outliers, which normalizes features and the target based on median and interquartile range 
before converting them into PyTorch tensors. The dataset is split into training ($70\%$), 
validation (15$\%$), and test (15$\%$) sets, with DataLoaders employed for batching (batch size = 128)
 and parallel loading. The core model, TransformerRegressor, comprises several key components: 
 an input embedding layer that linearly maps the 6-dimensional features to a 64-dimensional 
 latent space ($d_{model} = 64$); a sinusoidal positional encoding module that injects sequence 
 position information using sine and cosine functions with varying frequencies to address the 
 Transformer's lack of inherent positional awareness; a Transformer encoder consisting of 2 
 stacked encoder layers, each incorporating a multi-head self-attention mechanism with 4 heads, 
 a feed-forward network with a hidden dimension of 256 ($4\times d_{model}$), and dropout 
 regularization (rate = 0.1) to prevent overfitting; and an output head composed of two 
 fully connected layers (64$\rightarrow$32 with ReLU activation, then 32$\rightarrow$1) to 
 produce the final current prediction, with all high-dimensional parameters initialized 
 using Xavier uniform initialization. For training, the model is optimized using the 
 Adam optimizer (initial learning rate = 1e-4, weight decay = 1e-5) with mean squared 
 error (MSE) as the loss function, complemented by a ReduceLROnPlateau scheduler to 
 adjust the learning rate based on validation loss and gradient clipping (max norm = 1.0) 
 to stabilize training; early stopping (patience = 15) is implemented to halt training if 
 validation loss does not improve, with the best model saved based on minimum validation loss. 
 Evaluation on the test set involves inverse scaling of predictions and targets to their original 
 scales, followed by computing metrics including root mean squared error (RMSE), mean absolute 
 error (MAE), and $R^2$ score.

Additionally, before training, we attempted data cleaning on excessively discrete values to ensure data quality. Due to the inhomogeneity of device and material fabrication processes, some dark current data are anomalous, and data cleaning can enhance the predictive capability of the model. We also apply logarithmic transformation to the current, concentrating the data within the range of -14 to 5, which enhances model training efficiency and stability.


%
%
\section{Results and discussions }
\begin{figure}[H]  
 \centering
        \includegraphics[width=0.9\textwidth]{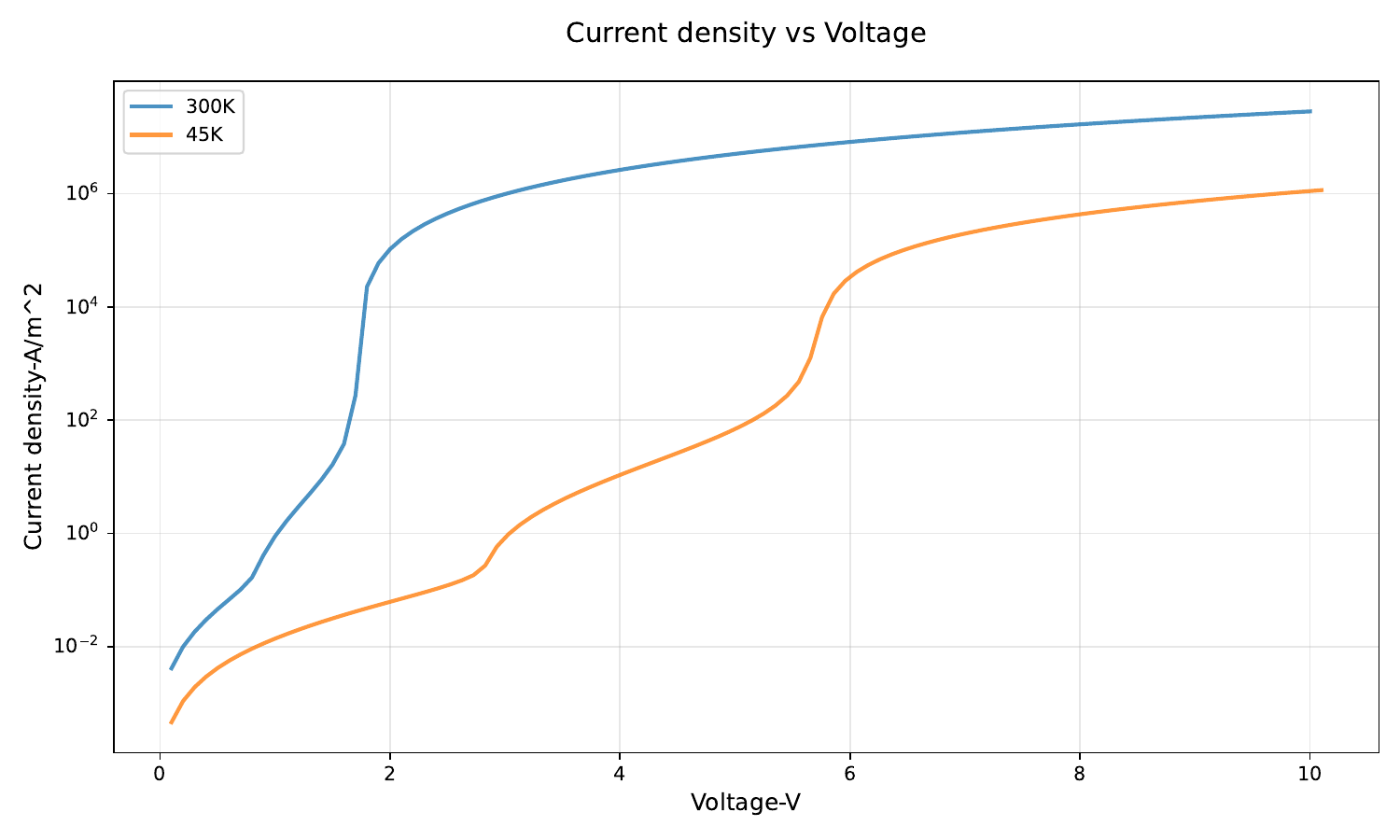}
 \caption{Theoretically predicted dark current curve. The unit of the ordinate is the logarithm of the dark current I to the base 10 (i.e., lg(I)), with the unit of $A/m^2$. The abscissa represents the applied voltage of the device, with the unit of volt (V).}
\label{fig6}
\end{figure}
Figure 6 presents the current surface density data calculated based on the SCLC theory. Within this voltage range, the data exhibit good overlap. The cross-sectional area of this type of device is relatively small, we find the predicted current surface density is similar to the dark current curve of Device 1 in Figure 4 in terms of both magnitude and profile by multiplying the current density by approximately \(10^{-9}\). This similarity confirms the validity of the proposed theoretical framework, the temperature promotes thermal excitation, leading to an increase in carrier concentration. Consequently, the current density curve shifts overall toward the upper-left direction.

\begin{figure}[htbp]
    \centering 
    
    \begin{subfigure}[b]{\textwidth} 
        \centering
        \includegraphics[width=0.8\textwidth]{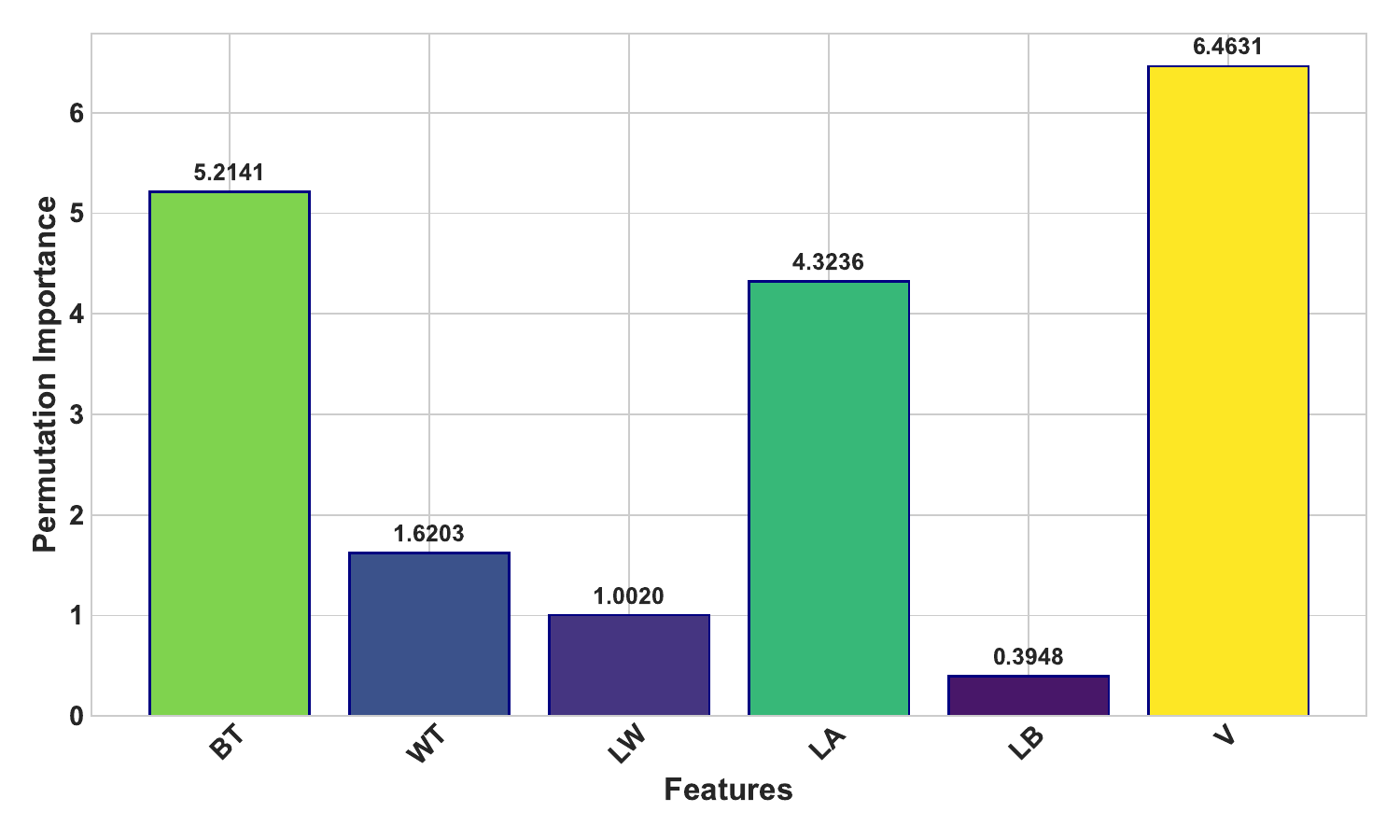} 
        \subcaption{Feature importance} 
        \label{fig:sub1} 
    \end{subfigure}
    \vspace{0.5cm} 
    
    \begin{subfigure}[b]{\textwidth}
        \centering
        \includegraphics[width=0.8\textwidth]{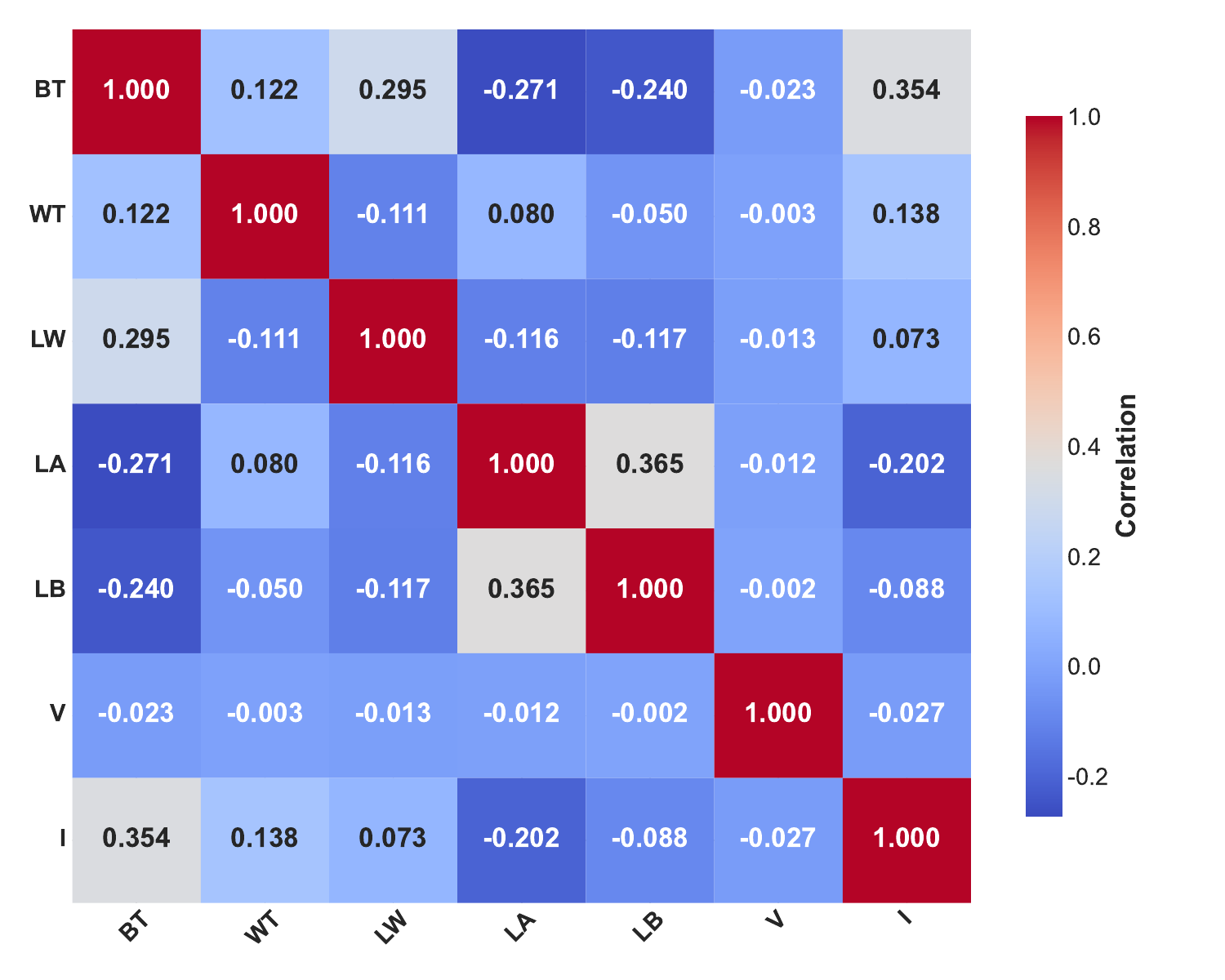}
        \subcaption{Correlation heatmap}
        \label{fig:sub2}
    \end{subfigure}
    \vspace{0.5cm} 
    
    
    
    \caption{Features of the dataset} 
    \label{fig:total} 
\end{figure}

The transformer architecture-based model achieved generally favorable performance, demonstrating a root mean 
square error (RMSE) of 2.1116 , mean absolute error (MAE) of 1.3565 , and coefficient of determination ($R^2$) of 0.5974 on 
the test dataset. Figure 7 illustrates the significance of device modeling parameters in dark current generation. Figure 7(a) reveals the degree of importance of each device with respect to dark current generation, while the bottom row of Figure 7(b) demonstrates the relationship between these devices and dark current generation. It can be concluded that BT and LA exert a relatively significant influence on dark current: specifically, a smaller BT value leads to a lower dark current, whereas a larger LA value results in a lower dark current. This provides valuable insights for engineering design.

\begin{figure}[htbp]
    \centering 
    
    \begin{subfigure}[b]{\textwidth} 
        \centering
        \includegraphics[width=1\textwidth]{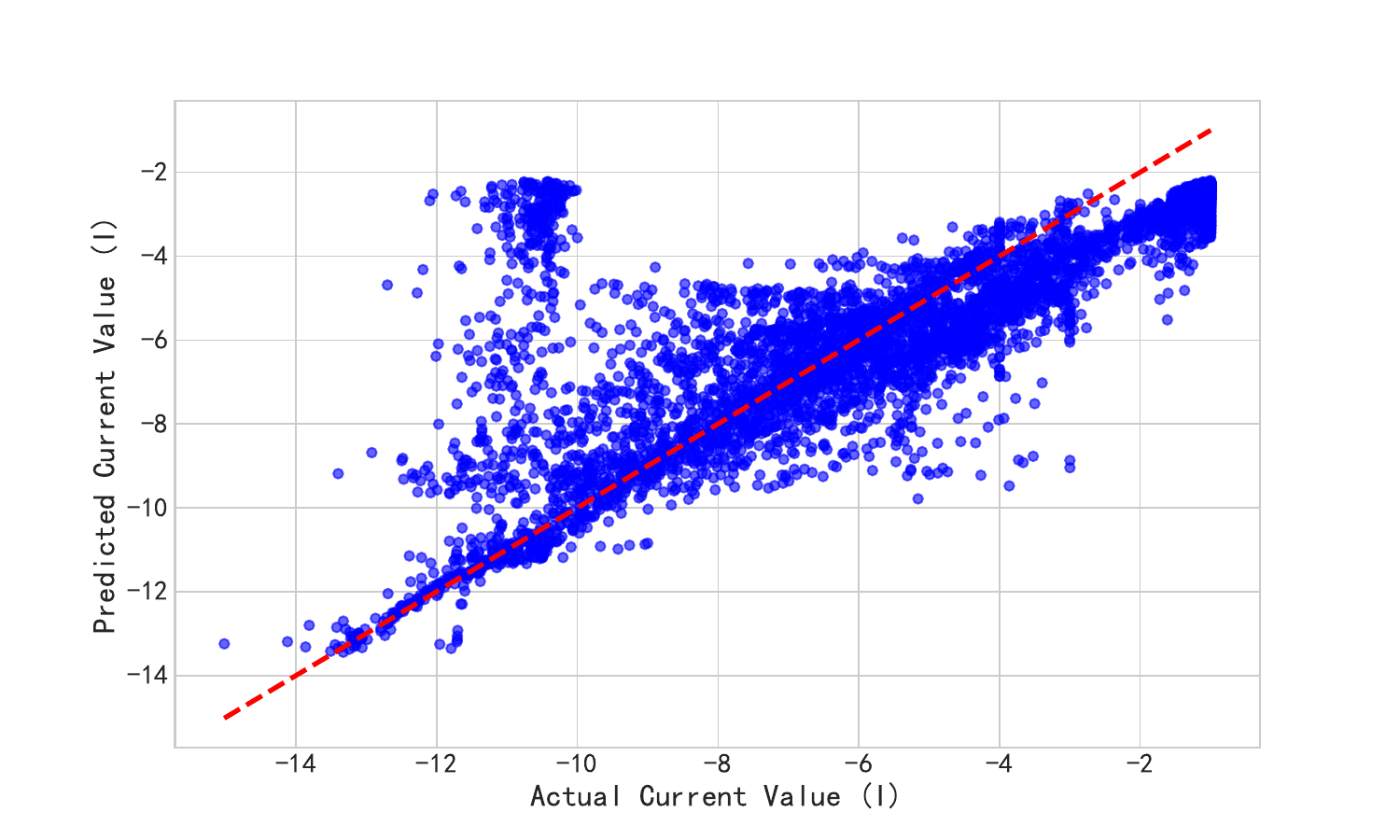}
        \subcaption{Prediction vs actual current.Both the horizontal and vertical coordinates are log(I)-transformed current, with the unit A.} 
        \label{fig:sub:upper} 
    \end{subfigure}
    
    \vspace{10pt} 
    
    \begin{subfigure}[b]{\textwidth}
        \centering
        \includegraphics[width=1\textwidth]{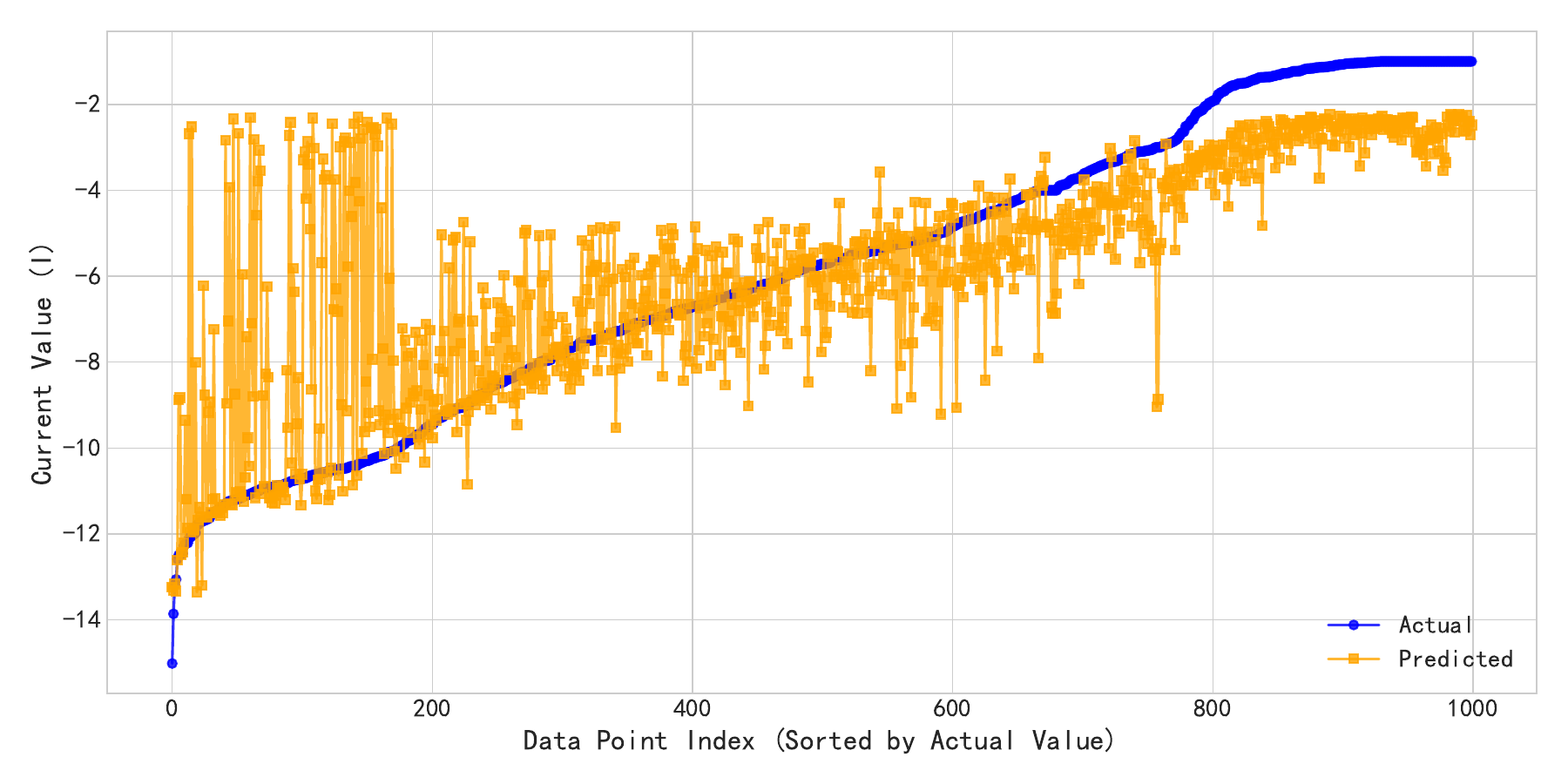}
        \subcaption{Actual vs predicted comparison.The horizontal axis represents the ordinal numbers of the data in the dataset, and the vertical axis represents the current processed by log(I) } 
        \label{fig:sub:lower} 
    \end{subfigure}
    
    \caption{Training consequence and model Predictions}
    \label{fig:combined} 
\end{figure}

Figure 8 presents the error distribution and overall performance of the model's prediction results. 
It can be clearly observed that the error deviates significantly within the current magnitude ranging 
from -12 to -10. This is attributed to the relatively low quality of some devices fabricated in the early stage, 
which inherently exhibit relatively high dark current; the inclusion of these data in the training process leads to an 
increased error. With the improvement of the manufacturing process and process standardization, the data quality will be 
further enhanced. Throughout the entire training process, the training and prediction data loss consistently decreases. However, overall, 
the data at the two ends of the voltage range are relatively difficult to predict. Such results suggest 
that there may be a phase transition of the transport mechanism associated with the electric field strength, which is also consistent with our transport physics model. This phenomenon can be attributed to two factors: on the one hand, measurement errors of small signals are 
relatively large; on the other hand, under low electric field conditions, irregular variations in current are induced by 
quantum fluctuations and multi-directional hopping conduction of electrons in the lattice with different probabilities\cite{22,23}.

\section{Conclusions}
The nonlinear behavior in the dark current of BIB devices originates from the inhomogeneous distribution of dopant atoms, while the different pathways of electronic hopping conduction lead to variations in the macroscopic dark current curve\cite{24}. Overall, the absorption layer makes the most significant contribution to charge carrier generation. Increasing its length may therefore suppress the magnitude of the dark current. Introducing trap states and material transitions can reduce the dark current, but this approach may induce oscillations in the dark current.
Through further experiments and measurements of characterization parameters in other dimensions of the device, more robust data-driven models can be developed. These models will enable predictions of the dark current for specifically designed devices and provide guidance for efforts aimed at dark current suppression.

\ack{This work is supported by the Research Funds of Hangzhou Institute for Advanced Study, UCAS.}





\end{document}